# ON THE ROLE OF THE KNUDSEN LAYER IN RAPID GRANULAR FLOWS

J. E. Galvin, C. M. Hrenya[*], and R. D. Wildman[1]

Department of Chemical and Biological Engineering, University of Colorado, Boulder, CO
    80309-0424, USA
[1]Wolfson School of Mechanical and Manufacturing Engineering, Loughborough University,
    Loughborough, Leicestershire, LE11 3TU, UK

**Abstract**

A combination of molecular-dynamics simulations, theoretical predictions, and previous experiments are used in a two-part study to determine the role of the Knudsen layer in rapid granular flows. First, a robust criterion for the identification of the thickness of the Knudsen layer is established: a rapid deterioration in Navier-Stokes-order prediction of the heat flux is found to occur in the Knudsen layer. For (experimental) systems in which heat flux measurements are not easily obtained, a rule-of-thumb for estimating the Knudsen layer thickness follows, namely that such effects are evident within 2.5 (local) mean free paths of a given boundary. Second, comparisons of simulation and experimental data with Navier-Stokes order theory are used to provide a measure as to when Knudsen layer effects become non-negligible. Specifically, predictions that do not account for the presence of a Knudsen layer appear reliable for Knudsen layers collectively composing up to 20% of the domain, whereas deterioration of such predictions becomes apparent when the domain is fully comprised of the Knudsen layer.

[*]corresponding author:  hrenya@colorado.edu

## 1. Introduction

Many practical granular flows are characterized by a lack of separation of length and time scales. Correspondingly, the appropriateness of continuum descriptions based on such an assumption has been a topic of ongoing debate (see, for example, Kadanoff, 1999, Goldhirsch, 2003). The analogous molecular system is a rarefied gas, which is characterized by a relatively high value of the Knudsen number, defined as $Kn = \lambda/L_{grad}$, where $\lambda$ is the mean free path and $L_{grad}$ is the length scale characterizing spatial variations in the hydrodynamic variables (Chapman & Cowling, 1970). Such systems, which lack a clear separation of scales, are characterized by large $\lambda$ and/or small $L_{grad}$, e.g., dilute flows, flow through microchannels, high Mach number (Ma) flows, low Reynolds number (Re) flows (where $Kn \propto Ma/Re$ for molecular gases). Examples of granular flows with similar characteristics include shallow flows down an inclined plane (Forterre & Pouliquen, 2001), supersonic flow past a wedge (Rericha et al., 2002), non-heaping grains at low gas pressure (Behringer et al., 2002), dilute flows past stationary objects (Wassgren et al., 2003), and the top layer of a vertically-vibrated, open system (Brey et al., 2001, Goldhirsch et al., 2004, Martin et al., 2006).

In *molecular systems*, the appropriate mathematical description of a gas depends on the value of Kn, since constitutive relations are derived based on series expansions about small values of Kn (Chapman & Cowling, 1970, Ferziger & Kaper, 1972). Generally speaking, for $Kn < 10^{-2}$ (continuum regime), Navier-Stokes order hydrodynamics coupled with a no-slip boundary conditions is appropriate. For $10^{-2} < Kn < 10^{-1}$ (slip regime), a slip (apparent) boundary condition is required to account for the presence of a non-negligible Knudsen layer[1] (Rosner & Papadopoulos, 1996). At even higher Kn, namely $10^{-1} <$ Kn $< 10$ (transition regime), higher-order terms in the Chapman-Enskog expansion (about Kn) are required, which increases the order of the governing equations and the corresponding boundary conditions. For example, the Burnett equations require second-order, slip boundary conditions to account for the Knudsen layer. Higher-order boundary conditions cannot be obtained solely from physical principles, and such descriptions remain an active area of research. Furthermore, solutions at the Burnett level are inherently unstable. One proposed solution to these problems is to consider relaxation in the system as a method for providing stability and closure (Jin & Slemrod, 2001). Despite this progress, however, for Kn $> 10$ (free molecular flow), a higher-order treatment for the domain interior coupled with a Knudsen layer at the boundary remains difficult to realize (Jin & Slemrod, 2001).

In the context of *granular flows*, the vast majority of theoretical contributions have focused on the continuum regime Navier-Stokes order hydrodynamics without modification to account for a Knudsen layer), whereas less attention has been focused on systems in which the separation of scales is not clear cut. At the high Kn limit (free molecular flows), Kumaran developed a model for shear flows with smooth particles (Kumaran, 1997) and rough particle-wall interactions (Kumaran, 2005), respectively. For transitional flows, Sela and Goldhirsch (1998) derived Burnett-order equations for dilute systems. Finally, several previous investigations have recognized the importance of the Knudsen layer at the open end of a vertically-vibrated system (Brey et al., 2001, Goldhirsch et al., 2004, Martin et al., 2006). Specifically, Martin et al. (2006) provided two estimates for the thickness of the Kn layer and Brey et al. (2001) developed a boundary condition to account for the Knudsen-layer effects. Both works are specific to the free surface of the vertically-vibrated system and do not consider the role of the Knudsen layer at the bottom (vibrating) boundary.

---

[1] Note that *all* wall-bounded flows contain a Knudsen layer adjacent to the wall in which the detailed nature of particle-wall collisions impacts transport as opposed to particle-particle collisions, which dictate behavior in the bulk interior. Similarly, systems with a free surface also display a Knudsen layer. For example, in a vibrated vessel with an open top, particles at the top surface are more likely to follow a parabolic trajectory than to engage in particle-particle collisions. Regardless of whether the boundary is open or closed, however, if the Knudsen layer is small relative to the characteristic system size, then an approach which does not incorporate the effects of the Knudsen layer is appropriate over the entire domain.

To build on the previous efforts, the aim of the current work is twofold: (i) to develop a robust method for determining the thickness of the Knudsen layer at a closed boundary, and (ii) to assess the appropriateness of a continuum-regime treatment (Navier-Stokes-order theory, coupled with no-slip conditions) for systems with various Knudsen layer widths (relative to system size). For the first purpose, molecular dynamics (MD) simulations of a quiescent (no mean motion) system with an imposed temperature gradient are utilized. This effort draws on previous work for the analogous molecular-gas (elastic) system investigated by Mackowski et al. (1999) and Pan et al. (2006). For the second purpose, a combination of MD simulations, (Navier-Stokes order) theory, and previous experiments are used. The results give rise to a general criterion for the identification of the Knudsen layer thickness based on heat flux measurements, and demonstrate under what conditions a Navier-Stokes order treatment that does not incorporate the effects of a Knudsen layer begins to deteriorate.

## 2.   Molecular Dynamics (MD) Simulations:  Computational Algorithm

This work employs three-dimensional, MD simulations of uniform particles to examine the effects of a Knudsen layer at the boundary. The particles are treated as inelastic, frictionless spheres. Particle collisions are assumed binary and instantaneous (i.e., hard-sphere assumption). As portrayed in Figure 1, the three-dimensional simulation domain is bounded on the left and right by motionless walls of constant, but not necessarily equal, granular temperature. The remaining four walls (top, bottom, front & back) are periodic. No body forces are present, and thus the system is characterized by zero mean flow. The simulation discussed herein has been described elsewhere; for additional details, the reader is referred to Dahl and Hrenya (2004) and Galvin, Dahl and Hrenya (2005).

As mentioned above, the simulation is bounded on the left and right side by walls of a set temperature ($T_{set}$). The boundary conditions employed in this effort are determined using a method for thermal walls presented by Cercignani (1987) and Poschel and Schwager (2005) and are slightly different than those employed in the simulations of the previously cited works (Dahl & Hrenya, 2004, Galvin et al., 2005). Particles colliding with one of the bounding walls are given a post-collisional velocity that is consistent with the $T_{set}$ of the wall with which they collided. Specifically, the post-collisional components of particle velocity ($c_{post}$) that are parallel to the wall (i.e., in the $y$ and $z$ directions) are determined as in the Box-Muller method for generating two tangential Gaussian distributions (Press et al., 1992):

$$c_{post,y} = \sqrt{-\frac{2T_{set}}{m_i} \ln(z_1)} \cdot \cos(2\pi z_2), \tag{1}$$

$$c_{post,z} = \sqrt{-\frac{2T_{set}}{m_i} \ln(z_3)} \cdot \sin(2\pi z_4), \tag{2}$$

where $z_1 - z_4$ are random numbers uniformly distributed in the interval [0, 1]. The post-collisional component of particle velocity normal to the wall (in the $x$ direction) is given by

$$c_{post,x} = \sqrt{-\frac{2T_{set}}{m_i} \ln(z_5)} \tag{3}$$

where $z_5$ is a random number again uniformly distributed in the interval [0, 1]. The sign (+ or −) of the pre-collisional component of particle velocity normal to the wall is reversed after collision.

The four remaining boundaries are standard periodic boundaries such that a particle crossing through one of these boundaries is returned through the opposing boundary with the same velocity and relative position.

The particles in the simulation are initially loaded onto a cubic lattice. The particles are then moved from their node positions by a small random displacement and any initial overlaps are resolved by making further random displacements in their position. As an estimate, the initial particle velocities are taken from a Maxwellian distribution that is consistent with their position along an assumed linear temperature gradient between the two set temperature walls. The actual temperature profile that develops between the two set temperature walls is non-linear, but the linear profile provides a basis for an initial condition.

The simulation proceeds in time via a hard-particle/overlap algorithm (Hopkins & Louge, 1991). In this method, the simulation progresses by making a series of small time steps during which the particles are moved along their linear trajectories. After each time step, any collisions are detected by searching for overlaps between particles or between a particle and a wall. Collisions are resolved using a hard-sphere model. For further details on the particle advancement algorithm, see Dahl and Hrenya (2004).

The input parameters for each simulation include $L_x$, $L_y$, and $L_z$, the length of the simulation domain in the $x$, $y$ and $z$ directions; $d$, the particle diameter; $m$, the mass of a particle; $\bar{v}$, the average solids volume fraction in the (entire) system; $e$, the coefficient of restitution; and $T_C$ and $T_H$, the set values of wall temperature located at $x/L_x = 0$ and $x/L_x = 1$, respectively. The dimensionless parameters that characterize the system are $\bar{v}$, $e$, $T_H/T_C$, $L_x/d$, $L_y/L_x$ and $L_z/L_x$. In this effort, a value of $L_x/d$ of 35 is used. This value was selected to ensure that each simulation was described by $1/Kn > 5$, where the "global" Knudsen number is defined as $Kn = \bar{\lambda}/L_x$ and $\bar{\lambda}$ $= d/(6\sqrt{2}\overline{\nu})$ is the spatially-averaged mean free path. Consequently, a given simulation will have a total number of particles ($N$) ranging from 4,000 to 12,000 depending on the values of the other dimensionless quantities ($\overline{\nu}$, $L_y/L_x$, $L_z/L_x$). In nearly all simulations the periodic domain lengths are set equal ($L_y = L_z$). For $L_x/d = 35$, a value of $L_y/L_x = L_z/L_x = 1$ is used to ensure that the collected data is not sensitive to further increases in the periodic domain length. As a result, the simulation domain is cubic ($L_x = L_y = L_z$) and the characteristic dimension is hereafter referred to as $L$. The remaining parameter space under investigation includes the temperature ratio ($T_H/T_C$) with set values of 1, 2 and 15, $\overline{\nu} = 0.025 - 0.15$ (Kn = 1.9 x 10$^{-1}$ – 3.2 x 10$^{-2}$) and $e = 0.8 - 1$. For convenience, the dimensional quantities $m$ and $L_x$ are set equal to 1.

The outputs from the simulation include lateral profiles of the solids volume fraction ($\nu$), granular temperature ($T$) and heat flux ($\boldsymbol{q}$), where the granular temperature is defined as $T = 1/3$ $<C^2>$ and $\boldsymbol{C}$ is the fluctuating velocity. Note that the dissipation rate ($\gamma$) is also collected in order to check whether the energy balance for the simulation is satisfied ($\nabla \cdot \mathbf{q} = -\gamma$; see Eq. 18), which serves as a verification of the simulation results. The check indicates that the energy balance is indeed satisfied; further details on calculating this quantity and the corresponding verification can be found in Hrenya, Galvin, and Wildman (2006).

Data collection begins once 10,000- 20,000 collisions per particle have occurred (depending on $\overline{\nu}$). To ensure that a statistical steady state has been achieved, the $x$-coordinate of the center-of-mass location of the particles and the granular temperature are monitored during the data collection period to make sure they change no more than 5% during the data collection interval. The reader is referred to Dahl and Hrenya (2004) for further details.

Several output quantities ($T$, $\boldsymbol{q}$, $\gamma$) are a function of the fluctuating velocity ($\boldsymbol{C}$), which in turn is defined relative to a local mass-average velocity. In this effort, a zero mass-average velocity is assumed in all calculations for the sake of computational efficiency. This assumption does not impact the results, as was established by Galvin, Dahl and Hrenya (2005). Accordingly, the fluctuating particle velocity equals the instantaneous particle velocity (*c*) and the two may be used interchangeably.

The spatial variation of the output quantities is determined by dividing the domain into thin rectangular boxes aligned in the direction parallel to the set-temperature walls. The width of each data collection strip ($\Delta x$) is set slightly wider than the particle diameter. Therefore, the current simulations include 30 data collection strips. The width of the collection strip (i.e. number of collection strips) was set so that the collected data does not change meaningfully with further resolution in the collection width (i.e., increased number of collection strips).

As mentioned earlier two types of data are reported in the simulation: hydrodynamic variables (*v*, *T*) and constitutive quantities (*q*). The solids volume fraction within each data collection strip is found by including only the volume of the particles that are within the data collection strip. The granular temperature within each data collection strip is found by including only the granular temperature of particles whose centers reside within the data collection strip at the instant of measurement. See Dahl and Hrenya (2004) for further details.

The total heat flux (*q*) consists of a kinetic component ($q_k$) and a collisional component ($q_c$). The heat flux is calculated following the methods used by Herbst, Müller and Zippelius (2005) in which the kinetic contribution of the heat flux is determined using

$$q_{k,a,strip} = \frac{1}{V_{strip}} \sum_{i=1}^{n_{strip}} \frac{1}{2} m C_{strip}^2 C_{a,strip} \ . \tag{4}$$

In this equation, $C_{a,strip}$ is the fluctuating particle velocity in that strip in the *a* direction (where *a* can be *x*, *y* or *z*), $C_{strip}^2 = \mathbf{C_{strip}} \cdot \mathbf{C_{strip}}$, $V_{strip}$ is the volume of the strip ($V_{strip} = L_z L_y \Delta x$) and $n_{strip}$ is the number of particles whose centers reside within the data collection strip. The collisional component of the heat flux is found by

$$q_{c,a,strip} = \frac{1}{2V_{strip}\Delta t} \sum_{coll_{strip}} \left( \Delta E_1 - \Delta E_2 \right) D k_a \qquad (5)$$

where $\Delta t$ is the elapsed time since data collection was initiated, $D$ is the distance between the particles centers, $k_a$ is the $a$ component of the unit vector pointing from the center of particle 1 toward the center of particle 2 ($a = x$, $y$, or $z$), $\Delta E_1$ is the change in energy of particle 1 due to a collision with particle 2 and $\Delta E_2$ is the change in energy of particle 2 due to a collision with particle 1. For a given particle, $\Delta E$ is defined as

$$\Delta E = \frac{1}{2}m\left(C_{post}{}^2 - C_{pre}{}^2\right) \qquad (6)$$

where the mass and fluctuation velocity are those quantities associated with the given particle. Note that collisional heat flux (Eq. 5) is found by summing only the heat flux of particles whose centers reside within the data collection strip during the collision (summation over $coll_{strip}$). In the event that the centers of the two colliding particles lay in different strips, the collisional heat flux is divided equally between the adjacent data collection strips in which the particle centers reside.

Unless otherwise noted, the data collection phase of each simulation comprises 50,000 collisions per particle during which 1,000,000 evenly spaced instantaneous measurements of solids volume fraction, granular temperature, and the kinetic components of the heat flux are made. The collisional components of the heat flux are evaluated as a summation over all collisions in the system during the data collection portion of the simulation. At the end of the simulation, the average of these measurements is calculated and reported. Measurements corresponding to the strips adjacent to the bounding walls are not reported since such measurements inherently include volume exclusion effects caused by the solid boundary (a particle cannot penetrate a solid boundary). For example, as the strip width approaches zero, the volume fraction at that strip will also approach zero. Such a width-sensitive measurement is not reflective of the actual hydrodynamic value, and is instead an inherent artifact of the averaging technique. Note that if the strip width is at least one particle diameter, which is the case for all systems examined here, such volume exclusion effects will only impact the strip adjacent to the boundary. Hence, the values obtained at strips adjacent to the boundaries are not being reported here. It is worth noting that because this wall value is not reported, the average solids volume fraction obtained by integrating the reported solids volume profile will not *appear* to be in perfect agreement with the average solids fraction initially set in the simulation, but the difference is small and the reported measurements are indeed accurate. It is also worthwhile to note that the observed temperature ratio will not match the set value of the temperature ratio (unless the ratio equals unity). The reason for the apparent mismatch can be traced to the thermal wall boundary condition, which only specifies the temperature of the outgoing particles (those which have collided with the wall) but says nothing about incoming particles (which will have a lower temperature). Since the measured temperature in the strip adjacent to the wall includes both types of particles (incoming and outgoing), the observed temperature will be lower than the set temperature. For simplicity, only the characteristic volume fraction and temperature ratio initially *set* in the simulation are reported.

## 3. Theoretical Predictions

To help elucidate the degree and impact of Knudsen layer effects, both constitutive quantities ($q_x$) and hydrodynamic variables ($v$ and $T$) obtained from MD simulations are compared to predictions obtained from two continuum theories for rapid, granular flows. In particular, the kinetic-theory-based predictions of Jenkins (1998) and Garzó and Dufty (1999) are considered. These theories are both targeted at uniform, inelastic, frictionless spheres engaging in instantaneous, binary collisions. Hence, the assumptions inherent in both theories mimic those of the MD simulations. Furthermore, both theories are of Navier-Stokes order (i.e., the constitutive relations are up to first order in spatial gradients). A key difference between the two approaches lies in the derivation process; the theory of Jenkins (1998) is based on an expansion about an elastic base case, while the theory of Garzó and Dufty (1999) is based on an expansion about a homogeneous cooling state. Correspondingly, the resulting constitutive relations take on a different form, the specifics of which are given below.

It is important to note that two distinct methods are used to compare the MD simulation data to the theoretical predictions. Specifically, the theoretical predictions for the constitutive quantity of interest ($q_x$) are obtained using MD simulation profiles as *inputs* to the theory, whereas the theoretical predictions for the hydrodynamic variables ($\nu$ and $T$) are obtained by solving the appropriate boundary-value problem (BVP). The type of information gleaned from each type of comparison is different in character, as is detailed in the following two sections.

### 3.1   Heat Flux:  Using MD data as inputs to theory

According to the two theories being considered here, the expressions for the heat flux for the bounded conduction problem considered here are:

*Jenkins* (1998)

$$q_x = -\left\{\frac{75}{64}\frac{m}{d^2\sqrt{\pi}}\sqrt{T}\left[\frac{1}{g_0} + \frac{24}{5}\nu + \frac{144}{25}\left(1 + \frac{32}{9\pi}\right)\nu^2 g_0\right]\right\}\frac{dT}{dx} \tag{7}$$

*Garzó and Dufty* (1999)

$$q_x = -\left\{\frac{75}{64}\frac{m}{d^2\sqrt{\pi}}\sqrt{T}\left[k_k^*\left(1+\frac{6}{5}\nu g_0\left(1+e\right)\right)+\frac{256\nu^2}{25\pi}g_0\left(1+e\right)\left(1+\frac{7}{32}c^*\right)\right]\right\}\frac{dT}{dx}$$

$$-\left\{\frac{25}{128}\frac{md\sqrt{\pi}T^{3/2}}{\nu}\left[\mu_k^*\left(1+\frac{6\nu}{5}g_0\left(1+e\right)\right)\right]\right\}\frac{dn}{dx} \qquad (8)$$

where

$$k_k^* = \frac{2}{3}\left(\nu_k^*-2\varsigma^{(0)*}\right)^{-1}\left[1+\frac{1}{2}\left(1+p^*\right)c^*+\frac{6\nu}{10}g_0\left(1+e\right)^2\right.$$

$$\left.\times\left\{2e-1+\left(\frac{1}{2}\left(1+e\right)-\frac{5}{3\left(1+e\right)}\right)c^*\right\}\right] \qquad (9)$$

$$\nu_k^* = \frac{1}{3}\left(1+e\right)g_0\left[1+\frac{33}{16}\left(1-e\right)+\frac{19-3e}{1024}c^*\right] \qquad (10)$$

$$\varsigma^{(0)*} = \frac{5}{12}g_0\left(1-e^2\right)\left(1+\frac{3}{32}c^*\right) \qquad (11)$$

$$p^* = 1+2\nu\left(1+e\right)g_0 \qquad (12)$$

$$c^* = 32\left(1-e\right)\left(1-2e^2\right)\left[81-17e+30e^2\left(1-e\right)\right]^{-1} \qquad (13)$$

$$\mu_k^* = 2\left(2\nu_k^*-3\varsigma^{(0)*}\right)^{-1}\left\{\left(1+n\frac{\partial\ln g_0}{\partial n}\right)\varsigma^{(0)*}k_k^*+\frac{p^*}{3}\left(1+n\frac{\partial\ln p^*}{\partial n}\right)c^*\right. \qquad (14)$$

$$\left.-\frac{12\nu}{15}g_0\left(1+\frac{1}{2}n\frac{\partial\ln g_0}{\partial n}\right)\left(1+e\right)\left[e\left(1-e\right)+\frac{1}{4}\left(\frac{4}{3}+e\left(1-e\right)\right)c^*\right]\right\} \qquad (15)$$

where *m* refers to the particle mass, *d* is the particle diameter, *e* is the restitution coefficient, $n = 6\nu/\left(\pi d^3\right)$ is the particle number density, and $g_0$ is the radial distribution function at contact, described here using the Carnahan and Starling (1969) expression:

$$g_o = \frac{2 - \nu}{2(1-\nu)^3}.$$ (16)

The constitutive relations for the heat flux are seen to depend on both constant material properties ($d$ and $m$ for both theories, as well as $e$ for the Garzó and Dufty (1999) theory) and hydrodynamic variables ($\nu$ and $T$). For the purposes of this work, it is important to note that the MD profiles for the hydrodynamic variables ($\nu$ and $T$) are used as *inputs* to the theoretical expressions for $q_x$ (rather than the $\nu$ and $T$ profiles obtained from the solution of the BVP, as described below). In this manner, any errors arising in the prediction of $\nu$ and $T$ are not propagated to the $q_x$ prediction, which thereby makes the comparison between the theoretical predictions for $q_x$ (Eq. 7 and 8) and those extracted directly from MD (Eq. 4 and 5) more clear-cut.

## 3.2   Concentration and Temperature:  Solution of boundary-value problem (BVP)

In addition to the heat flux comparison outlined above, comparisons between theory and MD simulations are also made for the hydrodynamic variables $\nu$ and $T$. Unlike the previous comparison, however, these profiles are determined via solution of the boundary-value problem describing the system. In this steady state, fully developed and quiescent flow, the conservation of mass is identically satisfied, and the conservation of momentum and the balance of granular energy take the following forms, respectively:

$$\frac{dP}{dx} = 0 \quad (\text{thus } P = \text{constant})$$ (17)

and

$$\frac{dq_x}{dx} = -\gamma \qquad (18)$$

where $P$ is the pressure and $\gamma$ is the dissipation rate of granular energy. The constitutive relations for $q_x$ are given by Eq. (7) and (8) for the two theories, and the corresponding pressures and dissipation rates for the system considered herein are given by:

*Jenkins (1998)*

$$P = \frac{6\nu}{\pi d^3} mT \left[ 1 + 4\nu g_0 \right]; \qquad (19)$$

$$\gamma = \frac{24\nu g_0}{d\sqrt{\pi}} \left( 1 - e \right) mn T^{3/2}; \qquad (20)$$

*Garzó and Dufty (1999)*

$$P = \frac{6\nu}{\pi d^3} mT \left[ 1 + 2\nu \left( 1 + e \right) g_0 \right]; \text{ and} \qquad (21)$$

$$\gamma = \frac{12\nu g_0}{d\sqrt{\pi}} \left( 1 - e^2 \right) \left( 1 + \frac{3}{32} c^* \right) mn T^{3/2}. \qquad (22)$$

The number of particles is set via an auxiliary integral equation for the average packing fraction (Arnarson & Jenkins, 2004, Wildman et al., 2006):

$$\bar{\nu} = \frac{1}{L_x} \int_0^{L_x} \nu \, dx \qquad (23)$$

which, to aid solution, is expressed as a differential equation

$$\frac{d\overline{\nu_{cum}}}{dx} = \frac{1}{L_x} \nu \left( x \right) \qquad (24)$$

where the subscript has been added to $\overline{\nu_{cum}}$ to indicate that it refers to the cumulative determination of average packing fraction (i.e., it is a function of $x$) as opposed to the overall packing fraction of the system, $\bar{\nu}$.

To solve for the four unknowns $\overline{v_{cum}}$, $v$, $T$ and $q_x$, the four first-order ordinary differential equations (Eq. 7 or 8, 17, 18, and 24) require boundary conditions for the granular temperature and the average packing fraction:

$$T\left(x=0\right)=T_C\,,\tag{25}$$

$$T\left(x=L_x\right)=T_H\,,\tag{26}$$

$$\overline{v_{cum}}\left(0\right)=0\,,\tag{27}$$

and

$$\overline{v_{cum}}\left(L_x\right)=\frac{\pi d^3 N}{6V}\,.\tag{28}$$

where $V = L_x L_y L_z$ is the system volume. The resulting system of equations is solved using the Matlab bvp4c boundary value solver (Kierzenka & Shampine, 2001) with the numerical inputs described in Section 2.

Recall that in the MD simulations, measurements corresponding to the strips adjacent to the bounding walls are disregarded. As a result, the MD system domain is effectively reduced and the parameters that characterize the resultant system (average volume fraction, temperature ratio of walls, etc.) are slightly different than the original input parameters. The characteristic domain length ($L_x$) for the reduced system is calculated based on the original domain length minus the width of three collection strips (one strip width arising from each bounding wall and an additional strip width since measurements are reported at the midpoint of a strip). Hence the values of $L_x/d$, $L_y/L_x$ and $L_z/L_x$, are all adjusted. The characteristic temperature ratio ($T_H/T_C$) is calculated based on the temperatures measured in the strips immediately next to those adjacent to the bounding walls. The average solids volume fraction is adjusted based on the solids volume that is actually present in the reduced domain. The adjusted values are then used to define the boundary value problem in order to obtain a one-to-one comparison between the reported MD system and the boundary value problem.

As will be demonstrated below, the heat flux comparison described in the previous section aids in pinpointing the boundary between the Knudsen layer and the bulk interior, while the solution of the BVP serves as an indicator of the appropriateness of Navier-Stokes order theory, coupled with no-slip conditions, for systems with Knudsen layers of various given thicknesses (relative to the system size).

## 4. Results and Discussion

As detailed above, the MD simulations were performed over a substantial parameter space. For the sake of brevity, the results presented herein are kept to a minimum and are representative of the complete set examined.

Before considering inelastic systems, it is helpful to first look at the more familiar case of a hard-sphere molecular gas ($e = 1$). Figures 2a and 2b display the MD results for the solids fraction and temperature profiles, respectively, for a case in which $\bar{v} = 0.05$ and $T_H / T_C = 2$. As is consistent with the equation of state (Eq. 19 or 21), the $v$ and $T$ profiles are inversely related. The characteristic "jump" in temperature at the wall, which is a hallmark of rarefied gases in the slip regime, is difficult to detect not only due to the scale of Figure 2b, but also since the value in the collection strip adjacent to the wall is not reported. Its presence, however, becomes apparent on a plot of the corresponding first derivative of temperature, as is shown in Figure 2c. First, the $dT/dx$ behavior in the interior of domain is distinct from that near the walls: an abrupt increase in the magnitude of $dT/dx$ is observed near both walls, which is consistent with the presence of a temperature jump. Second, note that $dT/dx$ displays a local minimum in the interior of the domain (at $x/L_x = 0.75$). As pointed out by Pan et al. (2006), this behavior is at odds with Fourier's law $q_x = -k \left( dT/dx \right)$ for heat conduction. More specifically, according to the energy balance for elastic systems ($e = 1$), the heat flux takes on a constant value (i.e., $q_x = -k \left( dT/dx \right) = \text{constant}$). Furthermore, both experiments and theory show that $k > 0$ and $dk/dT > 0$, i.e., $k$ is a positive value that increases with temperature (Chapman & Cowling, 1970, Bird, 1994, Cercignani et al., 1994, Sone, 2002). Thus, as $T$ increases, $k$ also increases, and thus according to Fourier's law, $dT/dx$ must *decrease* to maintain $q_x$ at a constant value. In contrast to this Fourier-law-based expectation, however, it is observed that while $T$, and thus $k$, increases as the hot wall is approached (from left to right in Figure 2b), $dT/dx$ shows a remarkable *increase* for $x/L_x > 0.75$ (Figure 2c). This violation of Fourier's law indicates that the Navier-Stokes-order constitutive relation for heat flux is no longer appropriate near the hot wall, thereby indicating that Knudsen effects are playing a role.

The characteristics noted above, namely the abrupt shift in $dT/dx$ and the violation of Fourier's law, are telltale signs of the Knudsen layer in molecular gases. As described below, however, such techniques for identifying the Knudsen layer prove less useful for inelastic systems. Hence, another criterion is introduced here which is found to work well in both elastic and inelastic systems. Specifically, the presence of a Knudsen layer near both boundaries is apparent when comparing the heat flux obtained from MD simulations to that obtained from (Navier-Stokes-order) theory. This comparison is given in Figure 2d using models based on the analysis of both Jenkins (1998) and Garzó and Dufty (1999). As described in Section 3.1, the theoretical predictions for $q_x$ are obtained using the MD profiles of $\nu$ and $T$ as inputs to the theory, so mismatches cannot be traced to any potential errors in $\nu$ and $T$. Recall for this system ($e = 1$) the heat flux should be constant, which is indeed obtained from the MD simulations (figure not shown).  The resulting error in $q_x$ between MD simulations and both theories (Figure 2d) is only a few percent in the interior of the domain, while it quickly increases to ~10% error and beyond toward the boundaries.  As previous studies for molecular gases have shown that deviations from Fourier's law are quite small even in the presence of large thermal gradients (Ciccotti & Tenebaum, 1980, Mareschal et al., 1987, Clause & Mareschal, 1988, Santos & Garzó, 1995), this abrupt mismatch can be traced to the  presence of a Knudsen layer (as opposed to the presence of large gradients or, equivalently, large Kn).  Furthermore, it is worthwhile to note that Burnett-order corrections to a dilute (Chapman & Cowling, 1970) or dense (Goldhirsch, 2006) molecular gas make no contribution to the heat flux for a zero-mean-flow system, so errors are attributable to effects of at least super-Burnett order.  Hence, these results are consistent with the findings of Pan et al. (Pan et al., 2006), whose empirical fit of the Knudsen-layer heat flux obtained from direct-simulation Monte Carlo (DSMC) of a molecular gas is third-order in temperature derivatives, implying that the Knudsen effects are beyond Burnett-order corrections.

Although the abrupt changes in the profiles of $dT/dx$ and the error in $q_x$ predictions provide a clear demarcation between the Knudsen layer and bulk interior, such profiles are difficult to obtain experimentally (whereas extraction from MD simulations is straightforward).  Hence, an alternative measure of the Knudsen layer thickness based on the volume fraction profile is also proposed.  The vertical lines appearing in Figure 2 provide such a measure.  Specifically, these lines represent a reciprocal (local) Knudsen number, defined as:

$$\frac{1}{Kn_{wall}} = \frac{\ell_{wall}}{\lambda_{wall}} = \left(\frac{\ell}{\lambda}\right)_{wall}$$

where $\ell_{wall}$ is the distance between a given wall and a point interior to the domain (e.g., the vertical line) and $\lambda_{wall}$ is the mean free path defined in terms of the average solids fraction ($\bar{\nu}_{wall}$) between the wall and the distance $\ell_{wall}$. For three-dimensional systems, such as those considered here, $\lambda_{wall} = d / \left( 6 \bar{v}_{wall} \right)$. In Figure 2, values of $(\ell / \lambda)_{wall}$ = 2.5 and 5 are represented by the thin (black) lines and thick (blue) lines, respectively. Moreover, the solid lines represent $(\ell / \lambda)_{wall}$ values associated with the cold wall (left boundary), whereas the dashed lines are those associated with the hot wall (right boundary). As evidenced by subplots (c) and (d), a value of $(\ell / \lambda)_{wall}$ = 2.5 provides a good demarcation between the Knudsen layer and the interior of the flow since this marker coincides with the abrupt changes observed in the $dT / dx$ profile (Figure 2c) and the rapid deterioration of Fourier's law (Figure 2d).

The physical reasoning for the suggested Knudsen-layer cutoff of $(\ell / \lambda)_{wall}$ = 2.5 is possible via a consideration of what is occurring at the molecular level. For the separation-of-scales assumption to hold in the interior of the domain (bulk region), the probability that a particle travels from the boundary to the domain interior without engaging in a collision must be small. For the simplified case of a homogeneous system (i.e., a molecular gas in thermal equilibrium), the probability ($P$) that a molecule of any speed travels a distance $L_D$ without a collision occurring is approximated by (Chapman & Cowling, 1970)

$$P = \exp(-1.04 L_D / \lambda)$$

where $\lambda$ is the mean-free path of the molecule. Note that for a value of $L_D / \lambda$ = 2.5, the chance of a particle traveling a distance $L_D$ without engaging in a collision is about 7%. Thus, according to the demarcation proposed above, the bulk region "begins" at a distance from the wall at which roughly 90% of particles, having encountered the wall, have subsequently engaged in a collision. Correspondingly, the Knudsen layer thickness ($\ell_{bl}$), is proportional to the mean free path, or more specifically, $\ell_{bl} = 2.5 *\lambda_{wall}$. Hence, a smaller mean free path (denser system) results in a thinner Knudsen layer, which is consistent with others (Mackowski et al., 1999, Pan et al., 2006) who have noted that the Knudsen layer thickness is proportional to the global Knudsen number ($Kn = \bar{\lambda} / L$).

To illustrate the robustness of the proposed Knudsen layer criterion for elastic systems, Figures 3 and 4 portray profiles analogous to those in Figure 2, but for a different set of parameters. Specifically, Figure 3 displays results for an increased average concentration of $\bar{\nu} = 0.10$, while Figure 4 contains results for an increased temperature gradient, namely $T_H / T_C = 14$, also for $\bar{\nu} = 0.10$. For both cases, $(\ell / \lambda)_{wall} = 2.5$ is again seen to provide a good indicator of a sudden change in the $dT / dx$ profile (Figures 3a and 4a) as well as the error associated with Fourier's law (Figures 3b and 4b). The high-temperature-gradient case is particularly noteworthy, as the thicknesses of the two Knudsen layers are quite different from one another due to the large difference in concentrations (and hence mean free paths) at each boundary.

The ability to identify the Knudsen layer for *inelastic* systems using $(\ell / \lambda)_{wall} = 2.5$ is demonstrated in Figures 5 and 6, which contains profiles for a system analogous to those of Figure 2 ($\bar{\nu} = 0.05$ and $T_H / T_C = 2$) and Figure 4 ($\bar{\nu} = 0.10$ and $T_H / T_C = 14$), respectively, except that $e = 0.99$. Since dissipation is now present, the energy balance is no longer represented by $\nabla \cdot \mathbf{q} = 0$ ($q_x = $ constant). Correspondingly, the aforementioned litmus test for the validity of Fourier's law, namely that $dT / dx$ should not take on a local minimum, is no longer valid. Despite the absence of such a test, the other salient features associated with the Knudsen layer are present, namely the temperature jump as indicated by the distinct change in the temperature gradient ($dT / dx$) profiles (subplots c), and the pronounced deterioration of the heat flux predictions in the Knudsen layer (subplots d), where it should be noted that the theories have incorporated the effect of inelasticity. Similar features are also exhibited in Figure 7, in which $\bar{\nu} = 0.05$, $e = 0.9$, and $T_H / T_C = 1$. It is also worthwhile to note that the theoretical prediction for $q_x$ in the domain *interior* is poorer than expected for inelastic systems, with errors of about 30% for $e = 0.9$. The source of this error is distinct from those caused by the Knudsen layer, and is related to the need to incorporate higher-order effects in the hydrodynamic description. This topic is explored further in Hrenya, Galvin, and Wildman (2006), together with analysis of other constitutive quantities, such as those for stress and dissipation rate.

At this point, several comments on the three criteria introduced thus far to identify the thickness of the Knudsen layer are warranted. First, as demonstrated by a comparison of Figures 5c and 5d at $x/L_x \sim 0.2$, a change in the $\mathrm{d}T/\mathrm{d}x$ profile at the boundary of the Knudsen layer may be subtle, whereas analogous changes in the error associated with $q_x$ predictions are abrupt. Second, although errors associated with the other constitutive quantities (stress tensor and dissipation rate of granular energy) were also examined, they do not exhibit as clear a demarcation between the Knudsen layer and bulk interior, and thus are not shown for the sake of brevity (but are contained in (Hrenya et al., 2006) for the interested reader). Third, for systems with non-thermal boundaries, the $(\ell/\lambda)_{wall} = 2.5$ criterion may be less robust than the other two due to a possible dependence on the detailed nature of the boundaries. Collectively, these observations indicate that the most robust indicator of the Knudsen layer thickness is that associated with the error in the $q_x$ profiles, and that the $(\ell/\lambda)_{wall} = 2.5$ criterion should be used as a rule-of-thumb (for systems with non-thermal boundaries) if measurements of $q_x$ are unavailable.

As explained in Section 3, two different types of comparisons between MD simulations and theory are investigated in this work. Up until this point, predictions of the constitutive quantity $q_x$ were obtained using MD profiles of $\nu$ and $T$ as inputs to the theoretical expressions (Eq. 7 or 8), which was then compared to that extracted directly from MD simulations (Eq. 4 and 5). As established above, this comparison is useful in delineating the Knudsen layer. As exemplified below, comparing the $v$ and $T$ profiles obtained from solution of the BVP to the corresponding MD profiles provide a measure as to the appropriateness of Navier-Stokes order theories, coupled with no-slip boundary conditions, for a given thickness of the Knudsen layer. More specifically, Figure 8 provides BVP solutions along with MD data for a relatively dense (Figures 8a and 8b) and dilute (Figures 8c and 8d) system. The systems are characterized by $T_H$ / $T_C$ = 1, $e$ = 0.99, and $v$ = 0.15 (dense) and 0.025 (dilute), respectively. As expected, the Knudsen layers are much smaller in the dense system than in the dilute system. For the latter, the lines representing $(\ell /\lambda )_{wall}$ = 2.5 almost meet in the middle of the domain (whereas the lines representing $(\ell /\lambda )_{wall}$ = 5.0 actually "cross" one another), indicating that nearly the entire domain is made up of the Knudsen layers. As is evident from these plots, the comparison between MD and theoretical predictions is excellent for the dense system in which the combined width of the Knudsen layers take up a relatively small fraction, namely ~20% combined, of the system domain. For the dilute system, on the other hand, a noticeable mismatch between the MD and theory is observed. This mismatch illustrates the need for slip boundary conditions for inelastic systems. Such conditions, which are well established for rarefied gases (Mackowski et al., 1999), represent the value of the hydrodynamic variable extrapolated from the bulk interior to the boundary; i.e., it is an "apparent" value rather than a "true" value, with the difference being referred to as the "slip". It is worthwhile to note that such corrections to the true value are proportional to the thickness of the Knudsen layer or, equivalently, the mean free path. Thus, as the Knudsen layers become thinner, the apparent value approaches the true value at the boundary. Hence, the BVP solution obtained using true (simulation) values of the wall temperature illustrate the impact of the Knudsen layer, namely systems with relatively small

Knudsen layers do not require slip conditions for accurate predictions in the bulk, (Figures 8a and 8b), while the opposite is true for systems with larger Knudsen layers (Figures 8c and 8d).

The above comparison between Navier-Stokes predictions and simulation data provides evidence that such a description, coupled with no-slip conditions, becomes less appropriate as the Knudsen layers begin to dominate this system. This idea is further supported by revisiting the experimental data of Viswanathan et al. (2006) for a vibro-fluidized bed. Specifically, Figure 9 shows granular temperature profiles obtained using positron emission particle tracking (PEPT) for a bed of particles which was vibrated at the base and left open at the top. The radially-averaged quantities are shown as a function of vertical distance, in which the positive $z$ direction points opposite to gravity ($g$). Temperature measurements associated with two directions are shown separately, namely $T_r = C_r^2$ and $T_z = C_z^2$, along with their scalar counterpart $T = (T_r + T_x + T_z)/3$; the other horizontal component $T_x$ is not shown since it is similar to $T_r$, which is the more accurate of the two horizontal measurements. Experiments were conducted for relatively dense (low vibrational amplitude) and dilute (high vibrational amplitude) systems, as displayed in subplots (a) and (b), respectively. Specifically, results for dimensionless base velocities of $V^* = V/\sqrt{gd} = 0.74$ and 1.54 (Figures 10a and 10b, respectively) are displayed, where $V$ is the velocity of the vibrating base. A vertical line is again shown in the plots for $(\ell/\lambda)_{wall} = 2.5$ to estimate the Knudsen layer originating from the vibrating base; this line is determined based on the experimental measurements of solids fraction as reported Viswanathan et al. (2006). (A similar line denoting the Knudsen layer at the open end of the system is not included since the bed height in an open container is not well-defined; estimates for the Knudsen layer thickness at such an open boundary can be found in Pasini & Jenkins (2005).) Also plotted are the corresponding theoretical temperature ($T$) predictions of Jenkins (1998) obtained from the solution of a two-dimensional, boundary-value problem. The BVP associated with this system differs from that of the MD system examined earlier due to the increased dimensionality, the presence of a gravitational field, and the use of no-slip boundary conditions; details are available in Viswanathan et al. (2006). Similar to the experimental measurements, the two-dimensional profiles are averaged in the radial direction and thus plotted as function of $z$ only. For the denser system (Figure 9a), the Knudsen layer takes up roughly 20% of the domain and the model predictions for $T$ and corresponding PEPT data are seen to be in fairly good agreement throughout the domain. For the more dilute system (Figure 9b), a noticeable mismatch between model predictions and experimental data occurs in the Knudsen layer, which represents about 30% of the domain. As a further testament to the presence of Knudsen layer effects in the dilute system, the experimental data of Figure 10b clearly indicates that the axial component of the temperature ($T_z$) "jumps" to a higher value at the base, which is characteristic of the temperature profile in the Knudsen layer near a hot wall, or energy source (1999). A similar jump is not observed in the denser system (Figure 9a) due to the reduced size of the layer. Note that a temperature jump in the direction perpendicular to the vibrating plate ($T_r$) is not observed in either case since the vertically vibrated bed provides an (anisotropic) energy source primarily in the vertical direction. Hence, the model-data mismatch noted above for the dilute system may arise from either the lack of an "apparent slip" boundary condition targeted at systems with non-negligible Knudsen layers (Mackowski et al., 1999) and/or the use of an isotropic theory in an anisotropic region. However, because particle-particle collisions will reduce the level of anisotropy caused by the vibrating base, the observed anisotropy is also restricted to the Knudsen layer region (where particle-particle collisions play only a small role), and hence both potential causes of mismatch are tied to the presence of a non-negligible Knudsen layer.

### 5. Concluding Remarks

The thrust of this effort has been twofold: (i) to identify the Knudsen layer in granular flows, and (ii) to assess its impact on predictions obtained from a Navier-Stokes-order theory applied across the domain with standard (no-slip) boundary conditions. First, it has been shown that the standard approaches used for identification of the Knudsen layer in molecular gases are not as useful for granular systems, since the presence of dissipation (i) may mask the abrupt change in the $dT/dx$ profile and (ii) precludes the use of the criterion based on the violation of Fourier's law (as heat flux is no longer constant). These shortcomings are overcome via an examination of the percentage error between MD data and Navier-Stokes order predictions for the heat flux (in which MD profiles of $\nu$ and $T$ are used to evaluate the theoretical expression). When this comparison is performed, a substantial increase in the mismatch between simulation and theory is observed, which provides a clear cut demarcation between the Knudsen layer and the bulk interior for both elastic and inelastic systems. For systems in which heat flux measurements are not readily available (such as experimental systems), $\ell_{bl} = 2.5 * \lambda_{wall}$ provides a rule-of-thumb for Knudsen layer thickness; this quantity (for a molecular gas in thermal equilibrium) corresponds to a distance from the wall in which less than 10% of the particles traveling that distance have not undergone a collision with another particle.

Second, the impact of the Knudsen layer on $\nu$ and $T$ predictions obtained from Navier-Stokes theory with standard (no-slip) boundary conditions has been examined via a comparison with both MD simulations and previous experimental data (Viswanathan et al., 2006) The comparison is quite good when the combined Knudsen layers represent about 20% of the system domain, while deterioration of the theory is evident when the domain is made up entirely of the Knudsen layer. The level of mismatch for the latter case, however, is perhaps less than might be expected, suggesting that the domain of validity for Navier-Stokes theory is greater than that afforded by a strict interpretation of the separation-of-scale assumptions. Although outside of the scope of the current effort, it is anticipated that the domain of validity could be further increased via the use of apparent (or slip) boundary conditions as is the standard approach for molecular gases with increasing rarefication.

In addition to the impact of the Knudsen layer on theoretical predictions, its impact on MD analysis and experimental design is also noteworthy. First, in several previous works, the heat flux has been extracted directly from MD simulations, either for development of a correlation (Soto et al., 1999) or for comparison with theory (Shattuck et al., 1999, Herbst et al., 2005). As is illustrated in the current work, care should be taken to ensure that such heat flux data is not extracted in the Knudsen layer, as the heat flux in the Knudsen layer takes on much different values than that in the bulk region. Analogous cautions are also recommended for other constitutive quantities, namely stress and dissipation rate. Alternatively, when using MD simulations to study the bulk interior, boundary conditions which effectively eliminate the presence of the Knudsen layer may be helpful, as have been used previously in direct simulation Monte Carlo studies of elastic systems (e.g., Montanero et al., 1994, Mackowski et al., 1999). Second, for experiments designed as tests of Navier-Stokes order theory, the size of Knudsen Knudsen layer should be kept in check.

It is also worthwhile to note that much recent interest has been focused on the form of the heat flux law for granular materials, and particularly on the importance of the term proportional to the concentration gradient (Sela & Goldhirsch, 1998, Soto et al., 1999, Herbst et al., 2005, Martin et al., 2006). For the bounded conduction problem examined in this work, this term does not play an important role, as is evidenced by the consistent agreement between the theories of Jenkins (1998) and Garzó and Dufty (1999), the former of which only contains a term proportional to the temperature gradient and the latter of which contains both terms.

Finally, although the focus of the current effort has been on the Knudsen layer, an interesting tendency is revealed in the domain interior (bulk region). Namely, an unexpected level of mismatch between the theoretical predictions for the heat flux and the corresponding MD values is observed. This behavior, as well as that of the other constitutive quantities (stress and dissipation rate) in the domain interior, are further explored in a companion work by Hrenya, Galvin, and Wildman (2006).

**Acknowledgements**

The authors would like to express their gratitude to Harish Viswanathan for providing model predictions for the vibrating bed system. Funding for this work has been provided by the Engineering and Physical Sciences Research Council (Grants EP/D030676/1 and GR/R75694/01). J.E.G. and C.M.H. would also like to thank the National Science Foundation for partial funding support via an international supplement to Grant CTS-0318999. C. M. H. and R. D. W. are also grateful to the organizers and participants of the Granular Physics Workshop at the Kavli Institute of Theoretical Physics (supported in part by the National Science Foundation under grant number PHY99-07949), as numerous discussions at the workshop provided motivation for much of this work.

**Figure 1.** Schematic of granular flow system.

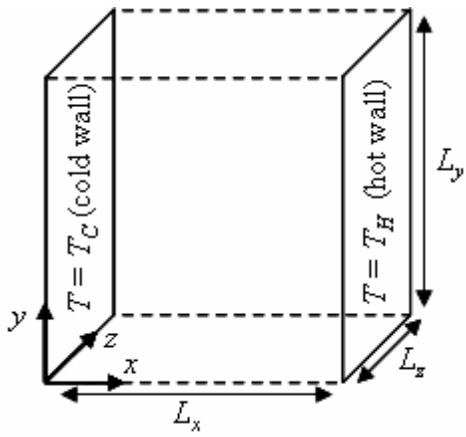

**Figure 2.** MD profiles of (a) solids volume fraction, (b) non-dimensional granular temperature and (c) the first derivative in granular temperature plus (d) the percent error in heat flux between MD simulations and theoretical predictions. MD simulations (circles); Jenkins (1998) predictions (thin solid line, red); Garzo & Dufty (1999) predictions (thick dotted line, black). Reciprocal local Knudsen numbers evaluated from the cold, left (solid) and hot, right (dash-dot) wall of 2.5 (thin, black) and 5.0 (thick, blue) are indicated by the vertical lines. Relevant parameters are $e = 1$, $\bar{\nu} = 0.05$ (1/Kn = 10.5), $T_H/T_C = 2$, $L/d = 35$. Data collection for this simulation involves 2,000,000 measurements over 100,000 collisions/particle.

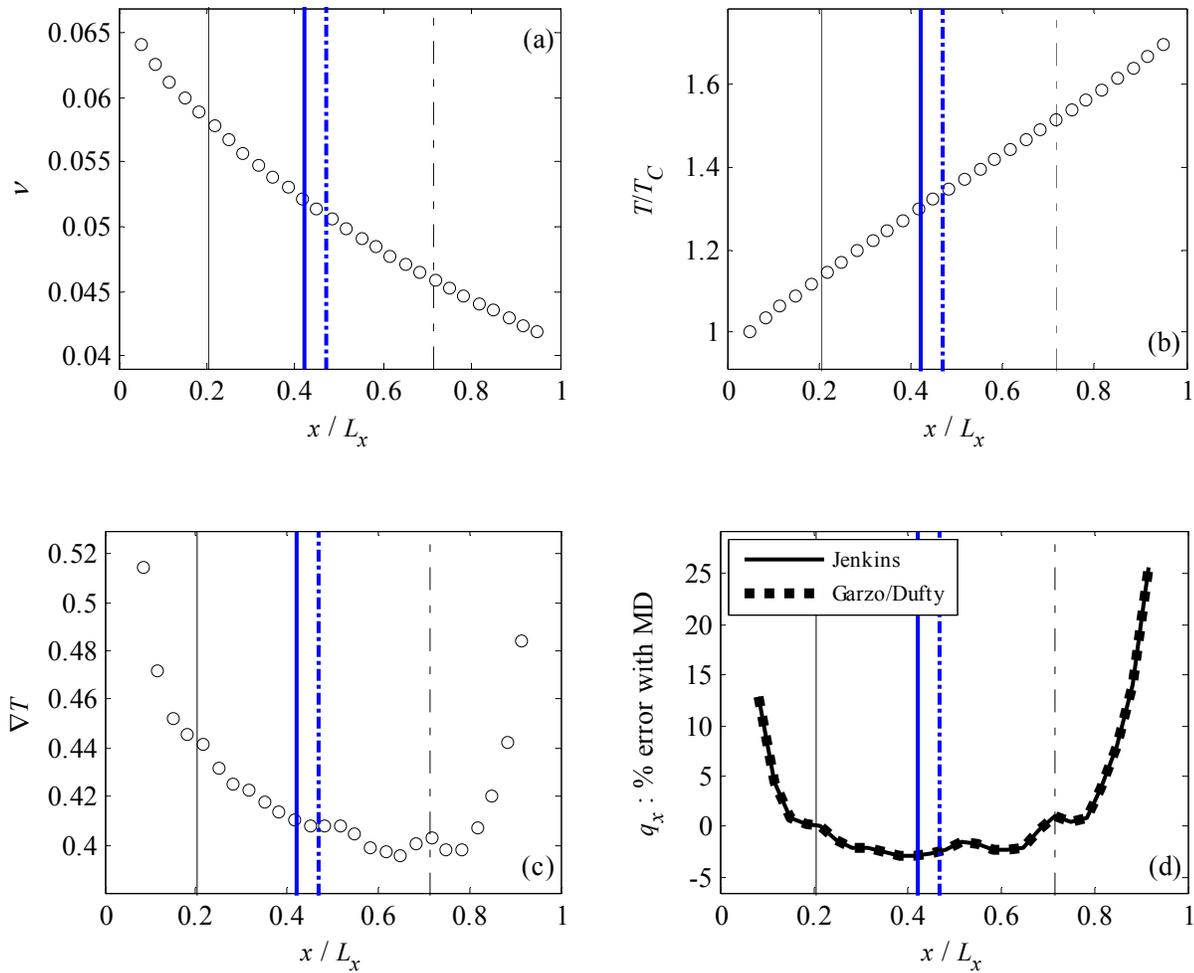

**Figure 3.** MD profiles of (a) solids volume fraction plus (b) the percent error in heat flux between MD simulations and theoretical predictions. Symbols and lines are as in Figure 2. Relevant parameters are $e = 1$, $\bar{v} = 0.1$ (1/Kn = 21.0), $T_H/T_C = 2$, $L/d = 35$. Data collection for this simulation involves 2,000,000 measurements over 100,000 collisions/particle.

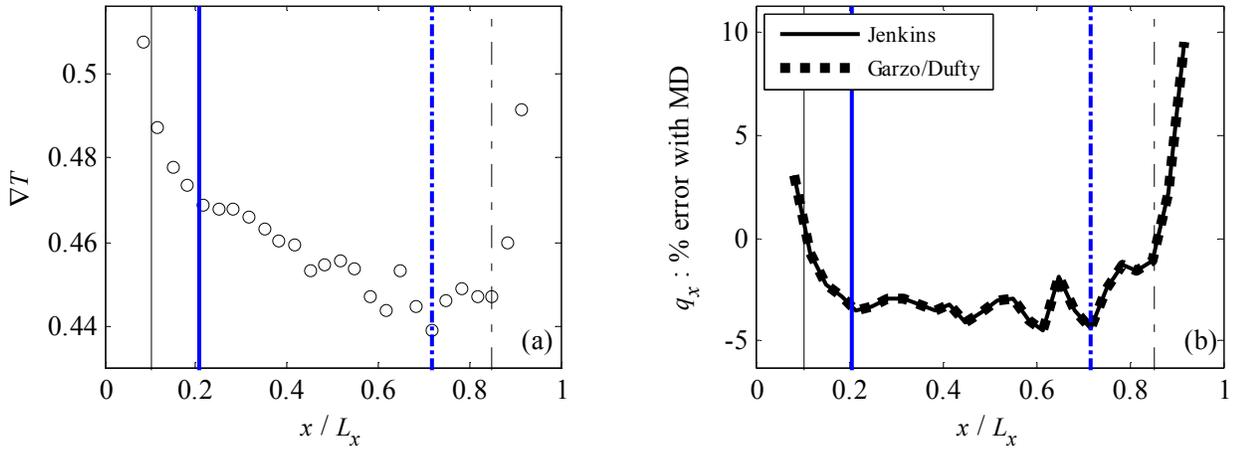

**Figure 4.** MD profiles of (a) solids volume fraction plus (b) the percent error in heat flux between MD simulations and theoretical predictions. Symbols and lines are as in Figure 2. Relevant parameters are $e = 1$, $\bar{\nu} = 0.1$ ($1/\mathrm{Kn} = 21.0$), $T_H/T_C = 14$, $L/d = 35$.

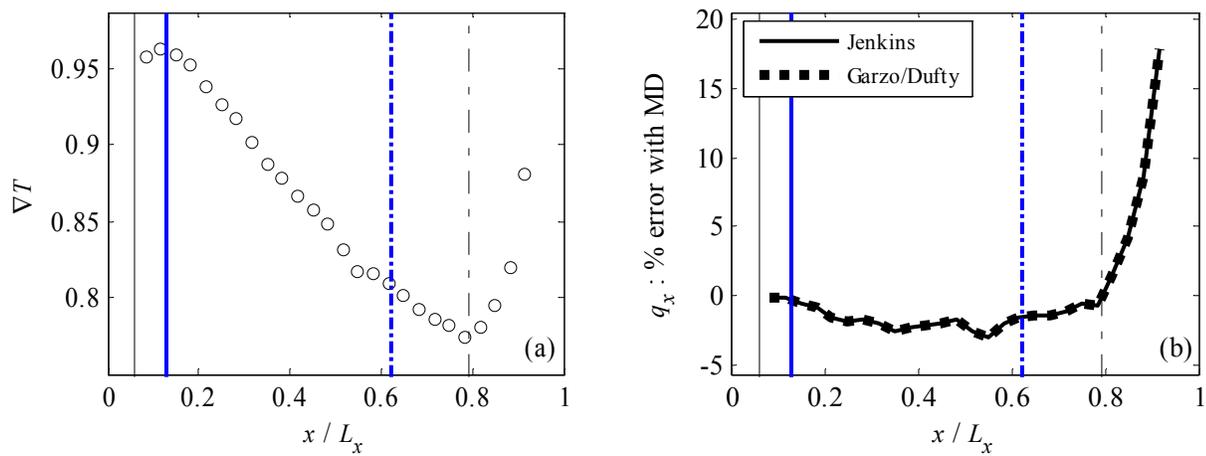

**Figure 5.** MD profiles of (a) solids volume fraction, (b) non-dimensional granular temperature and (c) the first derivative in granular temperature plus (d) the percent error in heat flux between MD simulations and theoretical predictions. Symbols and lines are as in Figure 2. Relevant parameters: $e = 0.99$, $\bar{\nu} = 0.05$ (1/Kn = 10.5), $T_H/T_C = 2$, $L/d = 35$.

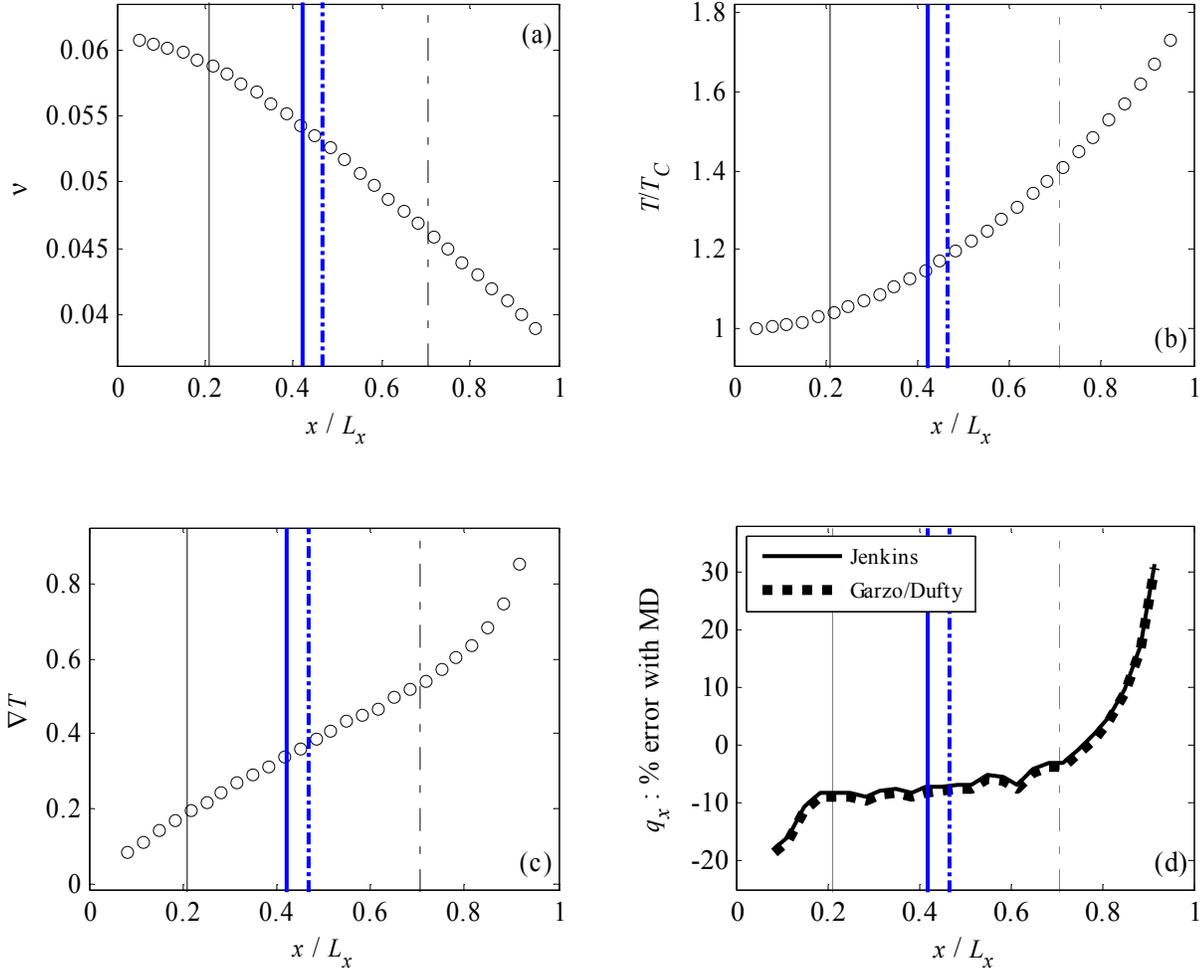

**Figure 6.** MD profiles of (a) solids volume fraction plus (b) the percent error in heat flux between MD simulations and theoretical predictions. Symbols and lines are as in Figure 2. Relevant parameters are $e = 0.99$, $\bar{\nu} = 0.1$ ($1/\text{Kn} = 21.0$), $T_H/T_C = 14$, $L/d = 35$.

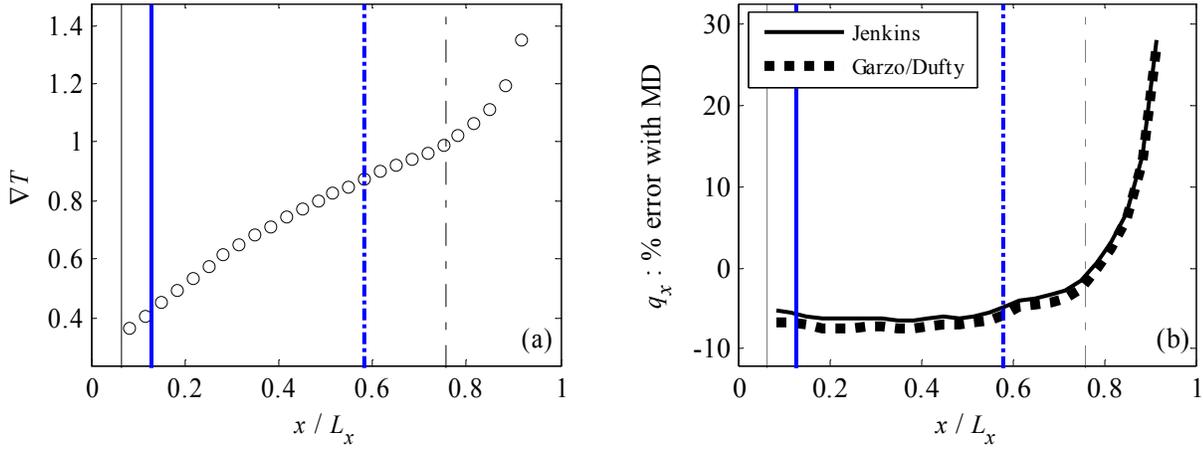

**Figure 7.** MD profiles of (a) solids volume fraction, (b) non-dimensional granular temperature and (c) the first derivative in granular temperature plus (d) the percent error in heat flux between MD simulations and theoretical predictions. Symbols and lines are as in Figure 2. Relevant parameters are $e = 0.9$, $\bar{\nu} = 0.05$ (1/Kn = 10.5), $T_H/T_C = 1$, $L/d = 35$. Data collection for this simulation involves 200,000 measurements over 10,000 collisions/particle.

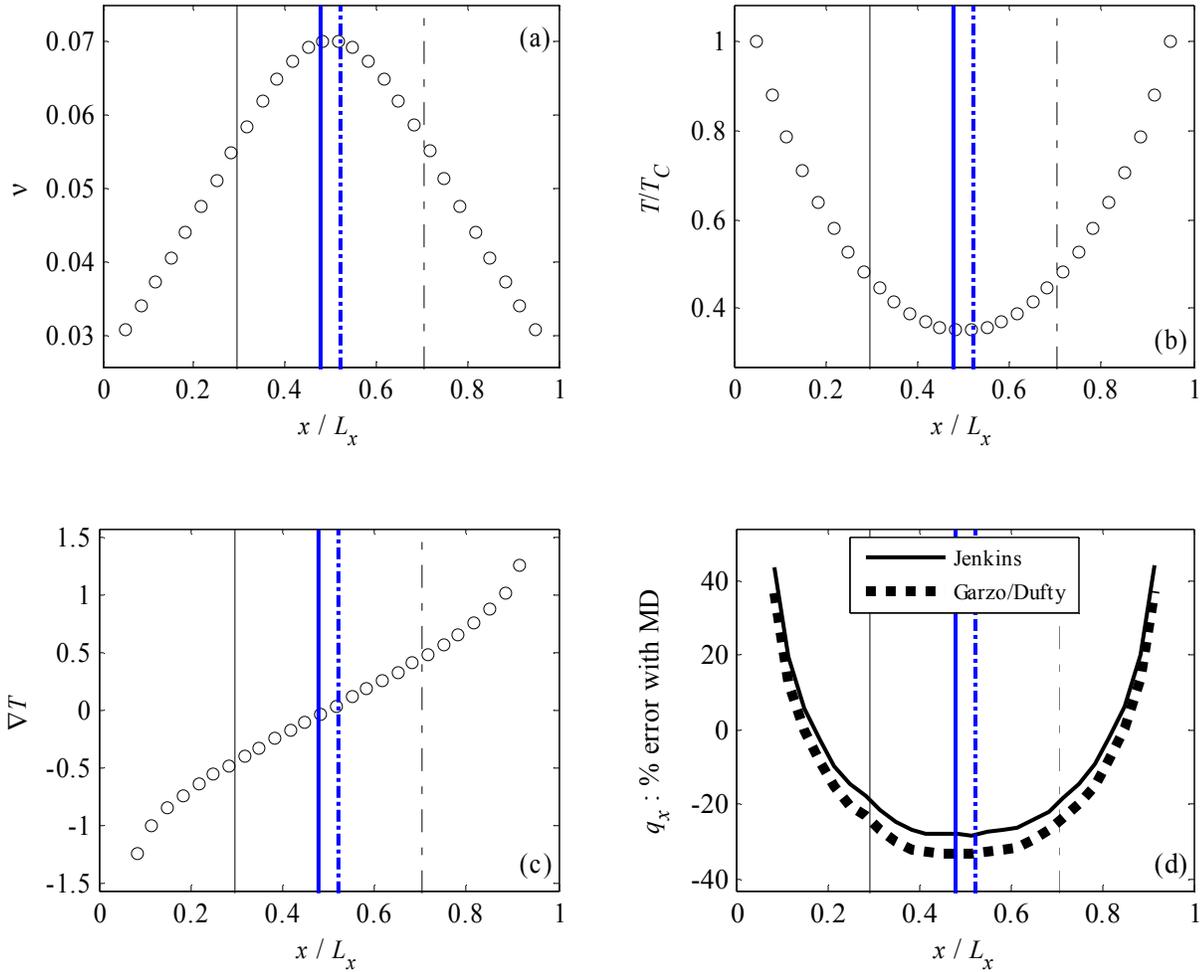

**Figure 8.** Profiles from BVP predictions based on Jenkins (1998) and Garzo & Dufty (1999) theories and from MD simulations of (a, c) solids volume fraction and (b, d) non-dimensional granular temperature for two different systems. Symbols and lines are as in Figure 2. Relevant parameters for (a) & (b) are $e = 0.99$, $\nu = 0.15$, $T_H/T_C = 1$, $L/d = 35$. Data collection for this simulation involves 200,000 measurements over 10,000 collisions/particle. Relevant parameters for (c) & (d) are $e = 0.99$, $\bar{\nu} = 0.025$, $T_H/T_C = 1$, $L_x/d = 35$ (1/Kn = 5.25), $L_z/d = L_y/d = 50$. Data collection for this simulation involves 500,000 measurements over 25,000 collisions/particle.

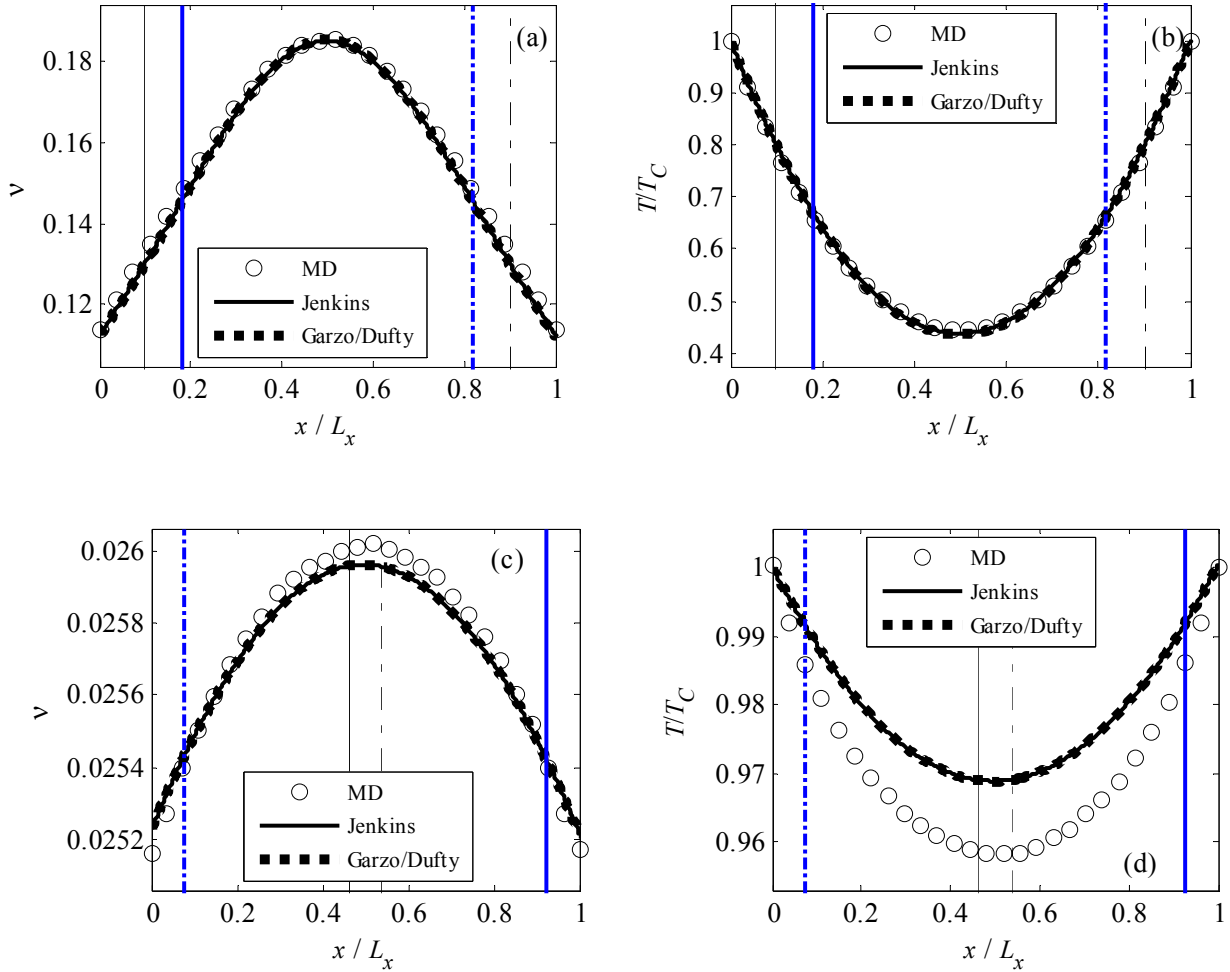

**Figure 9.** Temperature profiles in vibro-fluidized bed from BVP predictions of Viswanathan et al. (2006) based on Jenkins (1998) theory (red line) and from experimental PEPT data (individual symbols) of Martin et al. (2006). Reciprocal local Knudsen number of 2.5 evaluated from the vibrating, bottom boundary is indicated by vertical, solid black line. Relevant parameters are (a) $V^* = 0.74$ (dense) and (b) $V^* = 1.54$ (dilute)

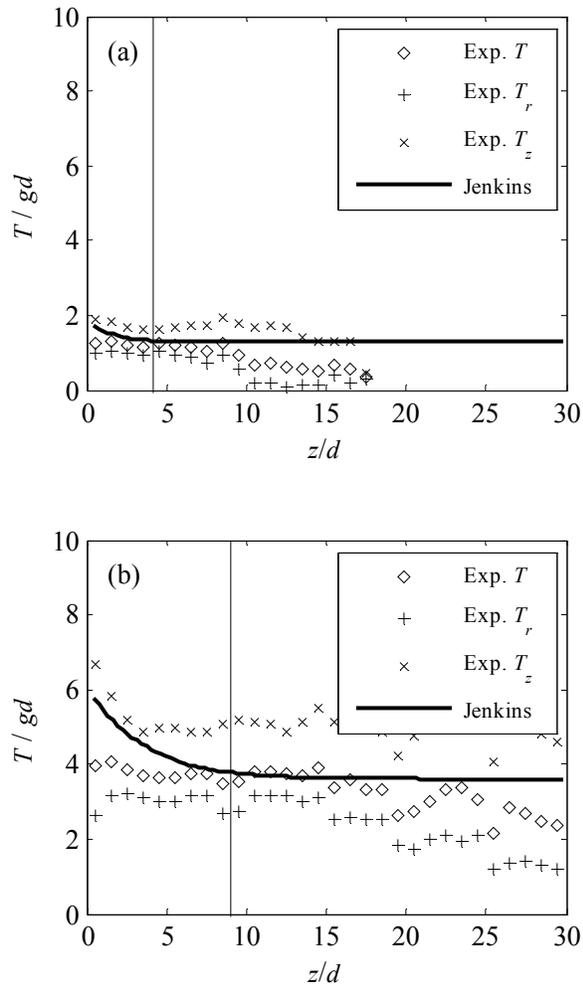